\documentclass[twocolumn]{aastex63}

\usepackage{ulem}
\usepackage{kotex}

\received{January 25, 2022}
\revised{April 29, 2022}
\accepted{May 6, 2022}
\submitjournal{ApJ}

\shorttitle{Circumnuclear medium in A1644-S}
\shortauthors{Baek et al.}
\graphicspath{{./}{figures/}}

\begin{document}

\title{Circumnuclear Medium around the Central AGN in a Cool-Core Cluster, Abell 1644-South}

\email{jhbaek1001@gmail.com, achung@yonsei.ac.kr}

\author[0000-0002-3744-6714]{Junhyun Baek}
\affiliation{Department of Astronomy, Yonsei University, 50 Yonsei-ro, Seodaemun-gu, Seoul, 03722, Korea}

\author[0000-0003-1440-8552]{Aeree Chung}
\affiliation{Department of Astronomy, Yonsei University, 50 Yonsei-ro, Seodaemun-gu, Seoul, 03722, Korea}

\author[0000-0002-3398-6916]{Alastair Edge}
\affiliation{Centre for Extragalactic Astronomy, Durham University, Durham DH1 3LE, UK}

\author[0000-0002-8310-2218]{Tom Rose}
\affiliation{Department of Physics and Astronomy, University of Waterloo, Waterloo, ON N2L 3G1, Canada}
\affiliation{Waterloo Centre for Astrophysics, Waterloo, ON N2L 3G1, Canada}

\author[0000-0002-1710-4442]{Jae-Woo Kim}
\affiliation{Korea Astronomy and Space Science Institute, 776 Daedeokdae-ro, Yuseong-gu, Daejeon, 34055, Korea}

\author[0000-0001-7003-8643]{Taehyun Jung}
\affiliation{Korea Astronomy and Space Science Institute, 776 Daedeokdae-ro, Yuseong-gu, Daejeon, 34055, Korea}




\begin{abstract}

\noindent We present the circumnuclear multi-phase gas properties of the brightest cluster galaxy (BCG) in the center of Abell 1644-South. A1644-S is the main cluster in a merging system, which is well known for X-ray hot gas sloshing in its core. The sharply peaked X-ray profile of A1644-S implies the presence of a strongly cooling gas core. In this study, we analyze ALMA $^{12}$CO~(1--0) data, JVLA \ion{H}{1} data, and KaVA 22~GHz data for the central region of A1644-S to probe the potential origin of the cool gas and its role in (re)powering the central active galactic nucleus (AGN). We find CO clumps distributed in an arc shape along the X-ray gas sloshing, which is suggestive of a connection between the cold gas and the hot intracluster medium (ICM). \ion{H}{1} and CN are detected in absorption against the AGN continuum emission. The absorption dip is observed at the systemic velocity of the BCG with an extended, redshifted tail. Based on the spatial and spectral configurations of the \ion{H}{1}, CN and CO gases, it is inferred that cool gas spirals into the core of the BCG, which is then fed to the central AGN. Indeed, our KaVA observation reveals a parsec-scale bipolar jet, implying that this AGN could have been (re)powered quite recently. Combining this, we suggest that some cold gas in A1644-S could have been formed from the cooling of the ICM, triggering the activity of the central AGN in the early development of a cool-core cluster.

\end{abstract}


\section{Introduction \label{sec:intro}}

\noindent A sharply peaked X-ray profile in the center of galaxy clusters indicates efficient cooling of the hot intracluster medium (ICM) by bremsstrahlung radiation (a.k.a. cool-core clusters). Gas may cool rapidly in the cluster center where the ICM density is the highest, which consequently induces the inward gas motion from the outer region to maintain hydrostatic equilibrium \citep[e.g.,][]{McNamara2007}. In the case of such cooling flows, it is predicted that gas can be funneled at a rate of up to several hundred solar masses per year \citep[e.g.,][]{Fabian1994, Fabian2012}. When this cool gas inflow is deposited in the center of the brightest cluster galaxy (BCG), nuclear activities such as star formation (SF) and/or the active galactic nucleus (AGN) can be enhanced. Indeed, the fraction of BCGs with detectable radio emission (i.e., the sign of AGN and/or SF) is found to be higher among cool-core clusters compared with non-cool-core clusters \citep[$>$~70\% vs. $<$~25\%;][]{Burns1990}. The cold gas present in the cluster center is, therefore, a crucial subject for better understanding the link between cooling flows and the nuclear activities of BCGs.

Single-dish observations using the IRAM\footnote{Institut de Radioastronomie Millim\'etrique}-30~m and the JCMT\footnote{James Clerk Maxwell Telescope}-14~m telescopes find that cold molecular gas associated with BCGs is not rare in the cool-core environment \citep[20~--~30\%;][]{Edge2001, Salome2003}. Recent high-resolution studies carried out using the Atacama Large Millimeter/submillimeter Array (ALMA) reveal a broad range of molecular gas morphology and kinematics of BCGs between two archetypes \citep[][and the references therein]{Russell2019, Olivares2019}: disk-dominated \citep[e.g., Hydra-A cluster;][]{Rose2019a} and filament-dominated \citep[e.g., Perseus cluster;][]{Lim2008}. The combination of the gas inflow formed by the ICM cooling and feedback from a powerful central AGN may yield a variety of cold molecular gas properties \citep[e.g., Abell 2597 by][]{Tremblay2018}.

ICM thermal condensation has been suggested as the main mechanism for the formation of cold molecular gas at the center of cool-core clusters \citep[e.g.,][]{Gaspari2013, Li2014}. However, most of the examples studied to date are the cases where cold molecular gas and well-developed AGN jets are observed simultaneously, and a complex cooling process in which AGN feedback is involved is likely at play \citep[e.g.,][]{Russell2019, Olivares2019}. In such environments, it is difficult to investigate the genuine role of ICM cooling and condensation as the origin of the molecular gas in BCGs. Therefore, to verify the hypothesis that condensed ICM can be the major source of the cool gas in BCGs, it is necessary to study the clusters where cool gas begins to form but extended jets around the central AGN are not present. These targets can also be an ideal laboratory to assess the impact of cooling gas on the activation of the cluster central AGN.

Abell 1644-South is one of the best examples of such environments. In this work, we thus probe the circumnuclear medium properties of Abell 1644-South to verify the connection between the ICM cooling and cool gas in the center of its BCG. We also investigate the role of cooling gas flow in the evolution of the BCG’s nuclear activities. This paper is organized as follows: in Section~\ref{sec:sample}, the environmental properties of Abell 1644-South are introduced. In Section~\ref{sec:obs}, we describe our multi-frequency data from various radio interferometers. In Section~\ref{sec:result}, we present the properties of circumnuclear medium and parsec-scale AGN jet of Abell 1644-South. In Section~\ref{sec:discussion}, we discuss the origin of circumnuclear cold gas and its role in the evolution of the BCG. We summarize this in Section~\ref{sec:summary}. In this study, we adopt a standard $\Lambda$CDM cosmology of $\Omega_{\rm m}$ = 0.315, $\Omega_{\rm \Lambda}$ = 0.685, and $H_{\rm 0}$ = 67.3 km~s$^{-1}$~Mpc$^{-1}$ \citep{Plank2014}.

\bigskip
\bigskip
\bigskip

\section{Abell~1644-South \label{sec:sample}}

\noindent Abell~1644 at $z$~=~0.047 \citep{Tustin2001} is a merging system of two subclusters, one in the north and one in the south (A1644-N and A1644-S hereinafter). A1644-S has more extended X-ray emission, implying that the southern cluster is the more massive of the two \citep[M$_{\rm 500}$~=~3.1~$\times$~10$^{14}$~$h^{-1}$~M$_{\odot}$;][]{Johnson2010}. It is also well known for its spiral-shaped X-ray sloshing, rotating in a counterclockwise direction from west to north \citep{Reiprich2004, Johnson2010, Lagana2010, Lagana2019}. \citet{Johnson2010} attributed this gas sloshing in A1644-S to the interaction with A1644-N about 700~Myr ago or even earlier. More recently, a third subcluster, A1644-N2, was identified by \citet{Monteiro2020} based on weak lensing analysis. Although N2 hosts no detectable X-ray emission, its dynamical mass from the weak lensing model is not negligible, being 84\% and 40\% of A1644-N and -S, respectively. Therefore, A1644-N2 may also have impacted the merging event in Abell~1644. Indeed, a numerical experiment by \citet{Doubrawa2020} showed that the ICM slosh could have formed 1.6~Gyr ago as a result of the interaction between S and N2.

Regardless of the detailed merging history, A1644-S must be where ICM cooling has recently started. Its hot gas morphology appears to be disturbed, yet it shows a steep decline in X-ray profile \citep{Reiprich2004, Johnson2010}. The interactions among the subclusters could have redistributed the ICM, sweeping away the thermal instability, which has been potentially induced by heating sources such as AGN jets launched in the past. In fact, no extended AGN jet at the cluster scale is present in the center of A1644-S, as confirmed in radio \citep{Reiprich2004}. All these factors might have led to the current ICM cooling and condensation, as indicated by the slope of the X-ray brightness. Therefore, A1644-S is a good example of an environment that is in the early stage of forming a cool gas flow, which is barely affected by the AGN jet activities.

\begin{table*}[ht]
	\centering
	\caption{The central position at individual wavelengths}
	\label{tab:tab1}
	\begin{tabular*}{0.86\textwidth}{lcccc}
		\hline
		 & RA & Dec. & Position offset & Data reference\\
		 & (hh:mm:ss.sss) & ($\pm$dd:mm:ss.ss) & (arcsec) & \\
		\hline
		r-band & 12:57:11.599 & $-$17:24:33.97 & -- & Pan-STARRS1 \citep{Flewelling2020} \\
		103~GHz cont. & 12:57:11.599 & $-$17:24:34.08 & $<$~0.15 & ALMA/2017.1.00629.S (P.I.: A. Edge) \\
		1.4~GHz cont. & 12:57:11.602 & $-$17:24:34.11 & $<$~0.15 & JVLA/15A-243 (P.I.: A. Edge) \\
		0.5-2.5~keV & 12:57:11.811 & $-$17:24:32.93 & 3.21 & Chandra/7922 (P.I.: D. Hudson) \\
		\hline
	\end{tabular*}
	\begin{flushleft}
	{\it Note.} The central positions have been measured by a single ellipse Gaussian fitting. They all coincide with the position of the peak at individual wavelengths within the fitting errors.\\
	\end{flushleft}
\end{table*}

In addition to hot gas, distinct phases of the medium in the core region of A1644-S have been investigated in different studies. Warm ionized gas traced by H$\alpha$ emission appears as filaments along the X-ray sloshing \citep{McDonald2010}. Based on the low ionization line ratios and line widths of the filaments, \citet{McDonald2012} suggested that ionized gas was formed through shocks from mixing layers rather than central AGN heating. Gas at lower temperatures has been detected as absorption lines against the central AGN continuum in \ion{H}{1} \citep{Glowacki2017} and CN~(1--0) hyperfine structures \citep{Rose2019b}. Both \ion{H}{1} and CN absorption lines consist of two components: one peaked near the BCG systemic velocity and the one redshifted with respect to the BCG center.

To better understand the connection between the hot ICM and the multi-phase medium that can feed the central AGN, we analyze the spatially resolved properties of circumnuclear gases and the AGN jet associated with the BCG of A1644-S. For this, we use the ALMA CO~(1--0) and CN~(1--0) data, and the Karl G. Jansky Very Large Array (JVLA) \ion{H}{1} data. In addition, we probe the parsec-scale jet properties of this target for the first time using the KVN and VERA Array (KaVA) 22~GHz continuum data. The A1644-S brightest cluster galaxy, PGC~44257, is located at RA:~12h~57m~11.599s, Dec.:~$-$17\,$^{\circ}$~24\,$\arcmin$~33.97\,$\arcsec$ (J2000). At the redshift of PGC~44257 \citep[$z$~=~0.047;][]{Tustin2001}, the physical scale is 0.96~kpc~arcsec$^{-1}$.

\section{Multi-frequency radio data \label{sec:obs}}

\subsection{ALMA data \label{subsec:alma}}

\noindent The ALMA $^{12}$CO~(1--0) data of our target were obtained with the 12~m array during Cycle~5 as part of the program 2017.1.00629.S (P.I.: A. Edge). This CO data was inspected by \citet{Rose2019b} with CN and SiO molecular lines for 18 BCGs, yet the spatially resolved CO properties of A1644-S are presented in this work for the first time. The observation was carried out in August 2018 with an on-source time of 46~min in Band~3. A single band of 1.875~GHz width was centered at 110.028~GHz with 488.281~kHz ($\sim$1.3~km~s$^{-1}$) channel separation, and multiple 2~GHz width bands were allocated to cover 95~--~99~GHz and 107~--~111~GHz with 15.625~MHz spectral resolution. J1229+0203 was observed for flux and bandpass calibrations, and J1245-1616 was used as the phase calibrator. 

In this study, we use the data products released by the ALMA Regional Center (ARC), which are processed as follows. A continuum map centered at 103~GHz was created using the 8~GHz width data in total. To spatially resolve the continuum source, uniform weighting and phase self-calibration were applied, yielding a synthesized beam of 1.5\,$\arcsec$~$\times$~1.0\,$\arcsec$. By performing a single ellipse Gaussian fitting, we find an unresolved source of 63~mJy at the optical center of the target in the continuum image. The morphology and the coordinate of this unresolved continuum source can be found in Figure~\ref{fig:ALMA_CO_maps2}(a) and Table~\ref{tab:tab1}, respectively.

The CO~(1--0) cube was produced using a single spectral window centered at 110.028~GHz. Briggs weighting of a robust 0.5 in the CASA-scale was adopted, yielding a synthesized beam of 2.0\,$\arcsec$~$\times$~1.4\,$\arcsec$, which is comparable to the CN~(1--0) data of \citet{Rose2019b}. The velocity of the spectral cube was smoothed to 8.9~km~s$^{-1}$ to match the request in the proposal. Using the ARC product, we created our own intensity, intensity-weighted velocity, and intensity-weighted velocity dispersion maps.

\subsection{JVLA data \label{subsec:vla}}

\noindent PGC~44257, the BCG of A1644-S, was observed at 1.4~GHz with the JVLA as part of the project 15A-243 (P.I.: A. Edge). Two observation runs with a 1.5~h on-source time each were performed in May 2015 using the BnA configuration. A single band of 16~MHz width was centered at 1.356~GHz with 31.25~kHz channel separation, which corresponds to $\sim$6.7~km~s$^{-1}$ at $z$~=~0.047. In addition, multiple 64~MHz width bands with 1~MHz resolution were allocated to cover 1~--~2~GHz for continuum detection. In both runs, 3C~286 was used as a flux and bandpass calibrator, and J1248-1958 was used for phase calibration. 

The data were processed using the standard CASA (5.6.2-3) pipeline. We manually flagged additional bad data and ran the pipeline multiple times. After the pipeline calibration, we applied phase self-calibration and merged the two datasets for imaging using {\tt tclean}.

\begin{figure*}[ht!]
    \epsscale{1.1}
    \centering
    \plotone{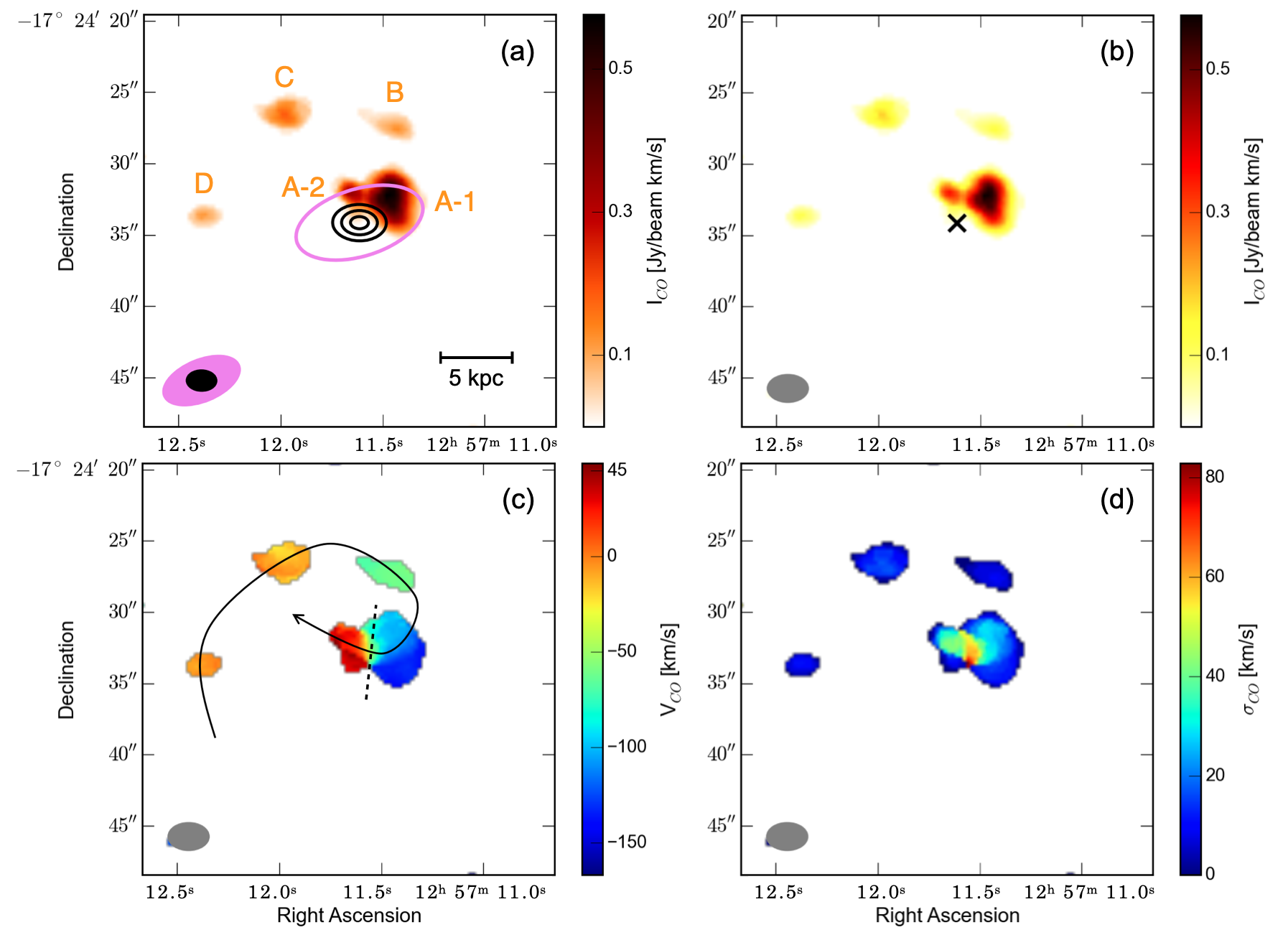}
    \caption{(a) JVLA 1.4~GHz continuum (magenta contour; 5$\sigma$ level) and ALMA 103~GHz continuum (black contours; 5, 10, 20$\sigma$ levels) are overlaid on the $^{12}$CO~(1--0) total intensity map of BCG in A1644-S. The synthesized beams of the JVLA continuum map (magenta solid ellipse) and the ALMA continuum map (black solid ellipse) are shown at the bottom left, and the physical scale bar of 5~kpc is shown at the bottom right. Both the JVLA and the ALMA continuums are unresolved and the central positions agree with each other ($\alpha_{2000}$=~12$^h$~57$^m$~11.60$^s$, $\delta_{2000}$=~$-$17\,$^{\circ}$~24\,$\arcmin$~34.1\,$\arcsec$).  (b) CO~(1--0) total intensity map is shown with a different color scheme to emphasize the brightness distribution of Cloud A. The synthesized beam of the CO~(1--0) (gray solid ellipse) is shown at the bottom left. The black cross symbol represents the position of the AGN continuum. None of the clouds were spatially located at this position. (c) and (d) CO~(1--0) intensity-weighted velocity field and intensity-weighted velocity dispersion, respectively. Here we adopt $v_{sys}$ of the BCG of 14,191~km~s$^{-1}$ (from stellar absorption lines in the optical barycentric standards; \citealt{Rose2019b}). Cloud~A, the largest structure can be divided into blue- and redshifted components (A-1 and A-2), as indicated by the black dashed line (see also (b)). Some emissions of A-1 and 2 are overlaid in the projected sky along the black dashed line, but they are not smoothly connected in velocity. The discontinuity of A-1 and 2 is also seen in Figure~\ref{fig:PVD}, the position-velocity diagram along the spline cut (black curved arrow). Starting from Cloud~A-1, the rest of the clouds are named as B, C, and D in the increasing order of line-of-sight velocities. \\}
    \label{fig:ALMA_CO_maps2}
\end{figure*}

We find an unresolved continuum source of 115~mJy at the optical center using a single ellipse Gaussian fitting. The JVLA 1.4~GHz continuum map and its central position can be found in Figure~\ref{fig:ALMA_CO_maps2}(a) and Table~\ref{tab:tab1}, respectively. We subtracted the continuum image from the cube to find \ion{H}{1} line emission. However, we do not find any detectable \ion{H}{1} emission although we probed a broad range of weighting and tapering parameters. Instead, we find a strong \ion{H}{1} absorption against its own continuum emission in the center. The final cube was created with uniform weighting to achieve the smallest possible synthesized beam (3.8\,$\arcsec$~$\times$~2.0\,$\arcsec$), based on which we constrained the size of the central continuum source. The spectral resolution was smoothed to 13~km~s$^{-1}$ for comparison with the CO spectrum and to achieve a high signal-to-noise ratio (S/N). The JVLA \ion{H}{1} data have a higher sensitivity and better angular resolution than the Australia Telescope Compact Array (ATCA) data presented by \citet{Glowacki2017}.

\subsection{KaVA data \label{subsec:kava}}

\begin{figure*}[ht!]
    \epsscale{1}
    \centering
    \plotone{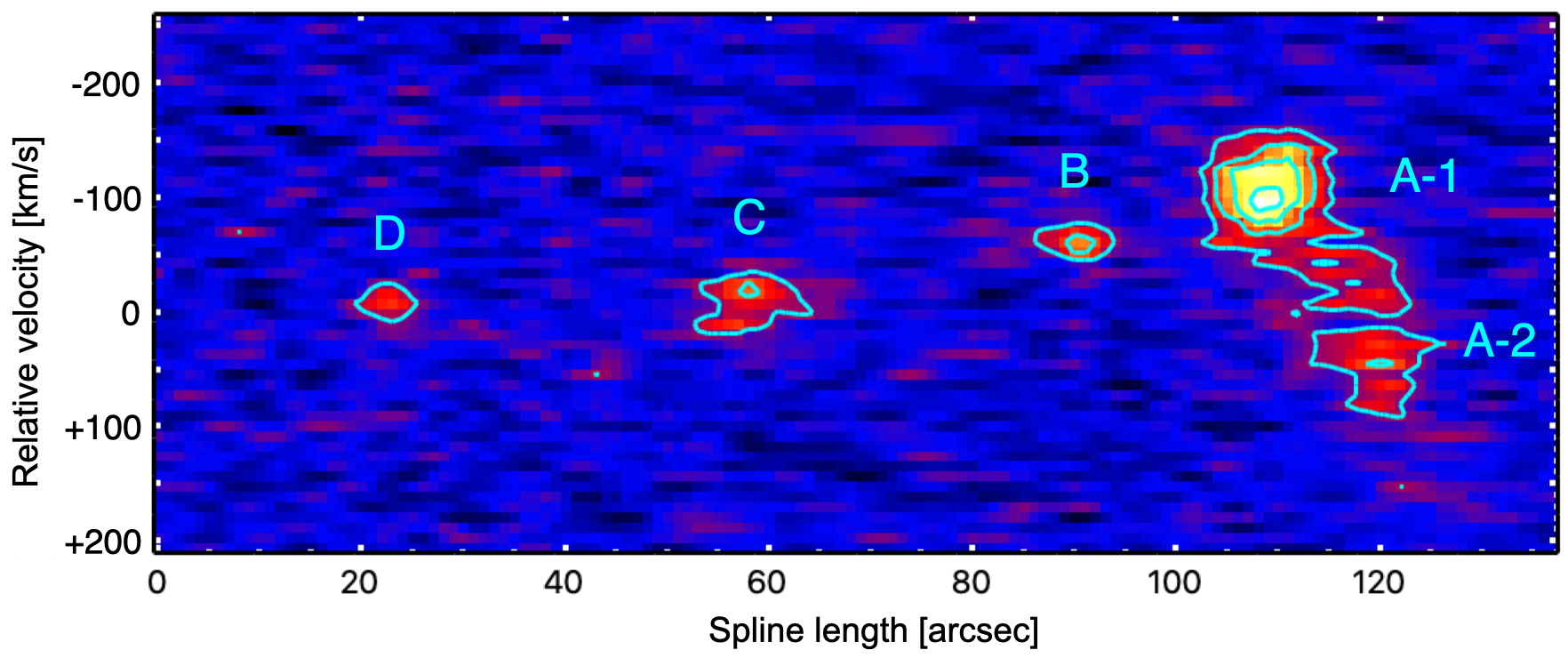}
    \caption{Position-velocity diagram (PVD) of CO~(1--0) made through a spline cut as indicated by a black curved arrow in Figure~\ref{fig:ALMA_CO_maps2}(c). The contour levels are 5$\sigma$~$\times$~2$^{n}$, n=0, 1, 2,..., where $\sigma$ is the RMS of the cube. While the radial velocities from Cloud~D to A-1 are smoothly decreasing, Cloud~A-2 does not seem to follow the overall velocity gradient. Although there is some diffuse gas between A-1 and A-2, it is highly clumpy, not smoothly connecting two A clouds. \\}
    \label{fig:PVD}
\end{figure*}

\noindent The KaVA 22~GHz continuum observation of PGC~44257 was carried out in April, 2020 (project ID: 20A-159; P.I.: J. Baek). KaVA is a combined array of the Korean VLBI Network (KVN) in South Korea and the VLBI Exploration of Radio Astrometry (VERA) in Japan. PGC~44257 is one of the targets showing the most significant variability in our KVN monitoring program of BCGs \citep{Rose2022}, and hence, its parsec-scale structure is of great interest. Data were obtained in a single sequence in the standard self-integration mode, observing four BCGs alternately for 20~min each to obtain as full a {\it uv}-coverage as possible for all targets. The total observation time was 11~h, including a few scans of a fringe finder (3C~345), yielding on-source time for each science target of 2.7~h. In this work, we focus on PGC~44257, and the full presentation of the entire sample will be published in a separate work.

The total of 512~MHz width band (two 256~MHz for each polarization) was configured with a central frequency of 21.4~GHz and a data recording rate of 1~Gbps. The data were correlated using the Daejeon correlator \citep{Lee2015a} and processed using the Astronomical Image Processing System (AIPS) package (31DEC15 version). Following the standard very long baseline interferometry (VLBI) calibration procedure, we first corrected the data for the ionospheric dispersive delay, voltage offsets in the samplers, instrumental delays, and parallactic angle variations. We then applied bandpass calibration and performed {\it a-priori} amplitude calibration using the gain curves and system temperatures. In the amplitude calibration, we multiplied the scaling factor of 1.3 to recover the quantization loss of the Daejeon correlator \citep{Lee2015b}. Finally, global fringe fitting was performed to correct for the residual fringe delays and rates of the target visibilities. Using the Difference Mapping (DIFMAP) software, we performed several iterations of clean, phase, and amplitude self-calibration until the solutions were found to have sufficient S/N as the root mean square (RMS) reached a desirable limit. We adopted a uniform weighting scheme to achieve the highest angular resolution for the given dataset (1.7~$\times$~1.0~mas$^{2}$). 

As a result, we find some extended emissions with a compact core in the imaging center. The absolute coordinate information of the KaVA~22 GHz data is lost over self-calibration as our observation was not designed for astrometry. However, it should be reasonable to infer that the pc-scale jet is located inside the unresolved JVLA and ALMA continuum emission.

\section{Results} \label{sec:result}

\subsection{Molecular gas properties} \label{subsec:CO_morphology}

\begin{figure*}[ht!]
    \epsscale{1}
    \plotone{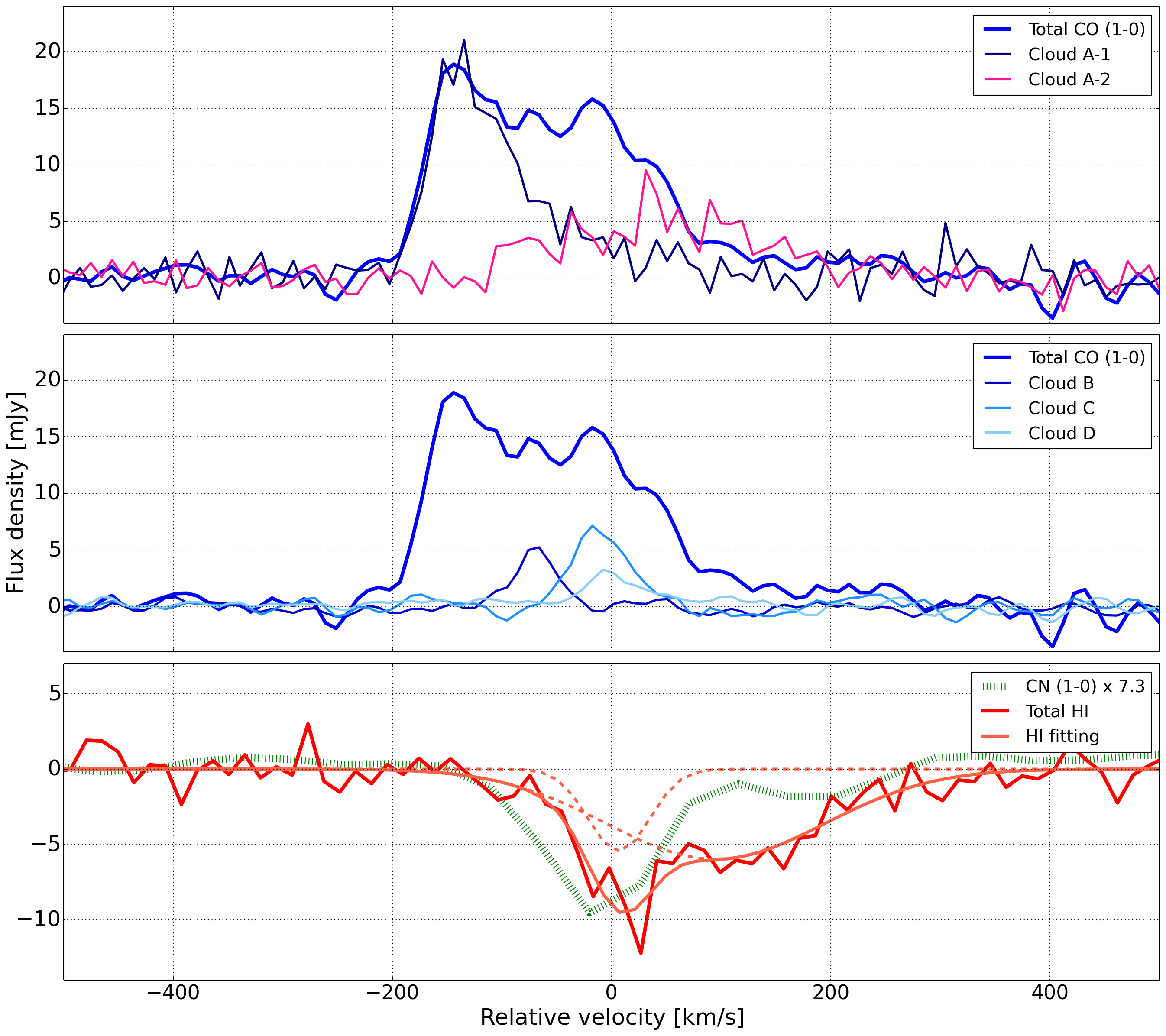}
    \caption{ALMA CO~(1--0) emission, CN~(1--0) absorption, and JVLA \ion{H}{1} absorption line profiles. The x-axis is the relative velocity with respect to the BCG system velocity (14,191~km~s$^{-1}$; \citealt{Rose2019b}). The integrated CO and \ion{H}{1} spectra are shown as blue and red bold lines, respectively. For CO gas, the flux densities of individual clouds are presented in blue colors, except Cloud~A-2 (pink color). From Cloud~D to A-1, the intensity peak gradually moves bluewards, which could imply an infalling gas stream from the rear side of the BCG. For \ion{H}{1} gas, the best-fitting Gaussian models of two components are presented as orange lines. The central narrow component and the redshifted broad component are suggestive of a rotating disk near the AGN and infalling gas clumps superposed in front of the BCG center, respectively. The CN absorption spectrum adopted from \citet{Rose2019b} is shown as a green line. The intensity of CN is amplified by a factor of 7.3 to match its peak with that of the \ion{H}{1} absorption. Similar to \ion{H}{1}, the CN absorption spectrum consists of a centrally peaked part and a redshifted component.\\}
    \label{fig:HICO_spectra}
\end{figure*}

\noindent From the ALMA observation, we find CO~(1--0) emission with the full width at half maximum (FWHM) of 250~km~s$^{-1}$. Similar to \citet{Rose2019b}, we do not find any clear CO absorption against the central continuum source but do see evidence of CN~(1--0) in absorption at low spectral resolution. In Figure~\ref{fig:ALMA_CO_maps2}, we show the total intensity map (Moment~0), the intensity-weighted velocity field (Moment~1), and the intensity-weighted velocity dispersion (Moment~2). The CO gas is resolved into multiple clouds, none of which are located at the position of the AGN (the center of each panel).

The most massive component, Cloud~A, can be decomposed into A-1 and A-2, as indicated by the dashed line in Figure~\ref{fig:ALMA_CO_maps2}(c). A steep velocity gradient is seen around $-$55~km~s$^{-1}$ with respect to the BCG system velocity (14,191~km~s$^{-1}$ from stellar absorption lines; \citealt{Rose2019b}), where the intensity peaks of A-1 and A-2 are blue- and redshifted by $-$140~km~s$^{-1}$ and $+$30~km~s$^{-1}$, respectively. In fact, these two structures do not appear to be very smoothly connected in the cube. As also seen in Figure~\ref{fig:PVD}, i.e., the position-velocity diagram made following the spline cut shown on Figure~\ref{fig:ALMA_CO_maps2}(c), there is some diffuse gas between A-1 and A-2 which are separated by $\sim$70~km~s$^{-1}$, but it seems highly clumpy which makes it hard to say that A-1 and A-2 are part of a single structure. This also yields a high velocity dispersion between A-1 and A-2, as shown in Figure~\ref{fig:ALMA_CO_maps2}(d).

The line-of-sight velocity changes smoothly from Cloud~D to A-1, forming an arc, as shown in Figure~\ref{fig:ALMA_CO_maps2}(c). Clouds~D and C have velocities similar to those of BCG, whereas Cloud~B is blueshifted and A-1 more so. The spatial and velocity distributions of these clouds highly suggest that they could be part of a single gas stream falling from the rear side of the BCG into its center. On the other hand, A-2, the redshifted component of Cloud~A is likely to be on the other side of the BCG, falling from the front to the BCG center. The relative positions and velocities of individual clouds can be also clearly seen in the position-velocity diagram in Figure~\ref{fig:PVD}.

\begin{figure}[t!]
    \epsscale{1.1}
    \plotone{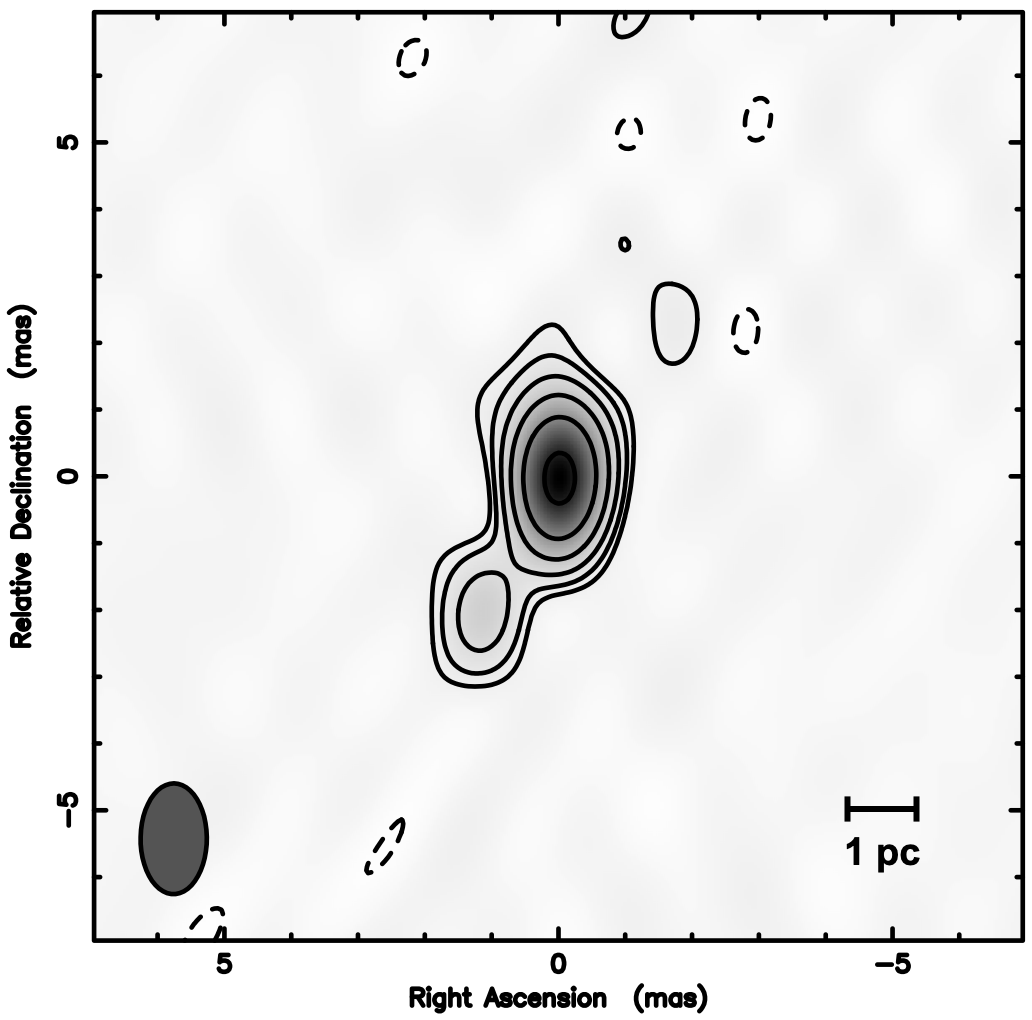}
    \caption{KaVA 22~GHz image of PGC~44257 (the BCG of A1644-S). The contour levels are (-3.6$\sigma$, 3.6$\sigma$)~$\times$~2$^{n}$, n=0, 1, 2,..., where $\sigma$ is the RMS of the image. The synthesized beam (1.7~$\times$~1.0~mas$^{2}$) is shown in the bottom left corner, and the bar in the bottom right corner indicates 1~pc at the distance of the target ($z$~=~0.047). The southeast extension around the bright core is clearly visible. A blob of 5.6-sigma is seen on the opposite side, suggesting the presence of a counter jet. The parsec-scale bipolar jet is likely to imply that the central AGN has recently been (re)powered.\\}
    \label{fig:KaVA_image}
\end{figure}

The flux density profiles of the individual clouds are shown in Figure~\ref{fig:HICO_spectra}. Adopting the Galactic $X_{\rm CO}$ of 2~$\times$~10$^{20}$~cm$^{-2}$~(K~km~s$^{-1}$)$^{-1}$ \citep{Bolatto2013}, the $\rm H_{2}$ mass of the clouds ranges from $\sim$10$^{10}~\rm M_{\odot}$ (Cloud~A) down to $\sim$10$^{8}~\rm M_{\odot}$ (Cloud~B$\sim$D). At the position of Cloud~B, UV emission has been detected \citep{McDonald2011}, indicating recent star formation. In Section~\ref{subsec:CO_origin}, we discuss the origin of this gas stream based on multi-wavelength data.

\subsection{Atomic hydrogen gas properties} \label{subsec:HI_absorption}

\noindent As shown in Figure \ref{fig:HICO_spectra}, our target is detected in \ion{H}{1} as absorption against its own central continuum emission. The maximum dip is found at 14,217~km~s$^{-1}$, which roughly agrees with the optical barycentric velocity of PGC~44257 measured using the stellar absorption lines \citep[14,191~km~s$^{-1}$;][]{Rose2019b}. The spectral shape is asymmetric owing to the extended tail on the high-velocity side. The absorption line can be fitted by two Gaussian components: the central narrow component (FWHM of 65.8~$\pm$~14.0~km~s$^{-1}$) at 6.6~$\pm$~4.6~km~s$^{-1}$ with a peak of $-$5.5~$\pm$~1.0~mJy and a redshifted broad component (FWHM of 234.0~$\pm$~19.8~km~s$^{-1}$) centered at 94.8~$\pm$~12.9~km~s$^{-1}$ with a peak of $-$6.0~$\pm$~0.5~mJy. Narrow absorption components in the center are commonly believed to trace a regularly rotating gas disk \citep[e.g.,][]{Morganti2018}, while broad components of a few hundred km~s$^{-1}$ can be formed by multiple gas clumps at different velocities along the line-of-sight \citep{Gereb2015}. The wide redshifted tail indicates that most of the cool atomic gas is receding from us. Hence, the \ion{H}{1} gas is likely to be an intervening medium between the continuum source and the observer, which is approaching to the BCG center.

We do not find any \ion{H}{1} emission associated with the target. Even if there is any, it might have been resolved out by the small synthesized beam. In Section \ref{subsec:HI_temp}, we estimate the spin temperature using a 3-sigma upper limit for \ion{H}{1} emission. We also constrain the size of the continuum source, for which radiation is absorbed by the atomic gas.

\subsection{Parsec-scale jet properties} \label{subsec:AGN_jet}

\noindent As shown in Figure~\ref{fig:KaVA_image}, the KaVA 22~GHz image reveals an extended feature to the southeast from a compact core. On the opposite side, a 5.6-sigma blob is present at almost the same projected distance as the peak of the southeast extension. These features are highly suggestive of a two-sided bipolar jet of 5.2~milli-arcsecond-size, which corresponds to 5.0~parsec at this redshift ($z$~=~0.047). The flux densities of the individual components were estimated by fitting a circular model to the visibilities. We measured 117.3~$\pm$~0.4~mJy from the core and 17.2~$\pm$~0.5 and 5.6~$\pm$~0.5~mJy from the SE and NW jets, respectively. This yields a total flux of 140~mJy, which agrees with that measured from the image.

The SE jet of the two is brighter by a factor of three, implying that the SE jet is the one launched toward us. In Section~\ref{subsec:jet_age}, we use this information to estimate the deprojected physical size of the bipolar jet, which also allows us to infer the approximate age of the jet.

\section{Discussion} \label{sec:discussion}

\subsection{The origin of the cold molecular gas} \label{subsec:CO_origin}

\begin{figure*}[ht!]
    \epsscale{1.1}
    \plotone{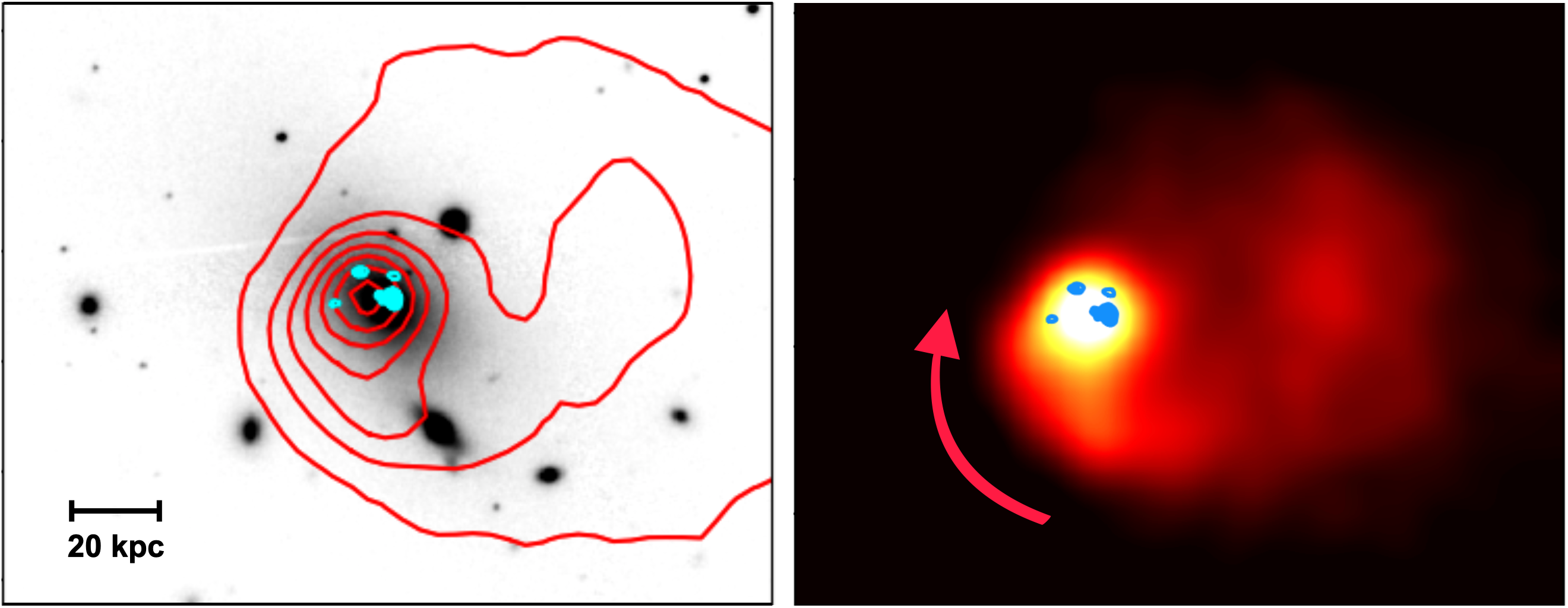}
    \caption{{\it Left}: Overlay of the Chandra X-ray (red contours) and ALMA CO~(1--0) (cyan shades) on the Pan-STARRS1 \citep{Flewelling2020} r-band grayscale image. The presented field is 165~kpc~$\times$~130~kpc in size and is centered on its BCG (PGC~44257). The bar on the bottom left indicates 20~kpc. Unlike the stellar morphology, which is relatively symmetric and unperturbed, both the X-ray emitting hot gas and the cold molecular gas traced by CO~(1--0) are asymmetrically distributed, forming a one-armed spiral shape. {\it Right}: The CO clumps are shown in blue shades on top of the X-ray image. The image scale is the same as the one on the left, but the center is shifted to the west by 0.3~arcmin in RA to better show the overall hot gas morphology. As indicated by the arched red arrow, the sloshing feature is suggestive of a gas flow spiraling into the BCG.\\}
    \label{fig:Overlay}
\end{figure*}

\noindent A1644-S, which shows signs of recent cluster-cluster merging and a steep X-ray profile simultaneously, is likely in the early stage of forming a cool gas flow in its center. There is no hint of kpc-scale AGN jets in the core region, and hence, it is inferred that this environment is not yet influenced by the thermal instability of the AGN jet. These characteristics make A1644-S an ideal laboratory to verify whether thermal condensation of the ICM can be a dominant source of cold gas deposition in the BCG.

In the left panel of Figure~\ref{fig:Overlay}, we present the overlay of the CO and X-ray intensity contours on the r-band optical image. Compared with the optical morphology of the BCG, both the CO and X-ray emissions appear to be asymmetrically distributed. The CO intensity peak of Cloud~A is offset from the optical center by 2.4~kpc to the northwest. Other clouds are scattered from the north, northeast to the east at 6.6~--~10.8~kpc radii from the optical center. The X-ray core of $\sim$13~kpc radius is symmetric, but its centroid is offset from the optical center by 3.1~kpc to the northeast (see Table~\ref{tab:tab1}). The positional offset between the BCG center and the X-ray peak of $\sim$3~kpc is not rare among the cool-core clusters \citep[e.g.,][]{Sanderson2009, Hamer2012, Birzan2012}. What makes A1644-S special is the asymmetry of the X-ray emission beyond $\sim$13~kpc from the peak due to the extent from the south to the southwest, that is, the sloshing feature.

Interestingly, the distribution of the CO clouds follows an arc shape. The spatial offset between the arc of the CO clouds and the stellar distribution suggests that the molecular gas is less likely to originate from the BCG. If the molecular gas is externally driven, a past wet-merging event can be a feasible scenario. To test this possibility, we inspected the optical residual after subtracting a coplanar ellipse model from the r-band image. However, we find no evidence of gravitational interactions between the BCG and other galaxies.

It is rather intriguing that the arc of CO gas clouds seems to be smoothly connected to the hot gas sloshing seen in the X-ray (see the right panel of Figure \ref{fig:Overlay}). This suggests a link between the two gas phases, cold molecular gas, and hot ICM. If the past merging of subclusters is responsible for the large-scale sloshing in the ICM, as suggested by \citet{Johnson2010} \citep[see also][]{Monteiro2020}, it could have induced the arc-shaped cold front by shock. This might have led to efficient gas cooling in the cluster core \citep[e.g.,][]{Roediger2011, ZuHone2013, Doubrawa2020}. Indeed, the cooling time of A1644-S is measured to be very short \citep[0.70$^{\rm +0.14}_{\rm -0.13}$~Gyr at the innermost X-ray core of 6.6~kpc radius;][]{Birzan2012}, and cold/dense gas condensation is known to be prevalent in systems where the central cooling time scale is less than 1~Gyr \citep[e.g.,][]{Pulido2018}. In the case of A1644-S, CO is mostly found within the innermost core region. Based on the multi-wavelength properties of A1644-S, it is therefore inferred that the CO clouds in its center are the product of the ICM cooling and condensation rather than the result of merging between galaxies.

\subsection{The spin temperature of atomic gas near the AGN} \label{subsec:HI_temp}

\noindent The \ion{H}{1} absorption in PGC~44257 was first reported by \citet{Glowacki2017} using the ATCA. Although the JVLA data used in this study achieved much better sensitivity and angular resolution than the previous observation (70.0\arcsec~$\times$~18.6\arcsec~for ATCA vs. 3.8\arcsec~$\times$~2.0\arcsec~for JVLA with uniform weighting), the \ion{H}{1} absorption profiles from the two facilities are consistent with each other within the observational uncertainties. The brightness of the unresolved continuum source in the center also agrees between the two measurements within 5\% of the flux. Therefore, the size of the continuum source can be better constrained by the smaller beam of the JVLA, and this confirms that there is no extended AGN jet associated with our target ($\lesssim$~3.6~kpc).

The gas temperature near the central AGN is an important physical quantity to be probed to understand the nature of the circumnuclear gas and its impact on galaxy evolution. To constrain the \ion{H}{1} spin temperature $T_{\rm S}$, both the \ion{H}{1} emission and absorption measurements are required in regions close enough that are presumably part of a single cloud. Unfortunately, neither in the previous ATCA observation of a larger synthesized beam nor in our further probe of the JVLA data with various imaging parameters such as weighting scheme and {\it uv}-tapering did we detect \ion{H}{1} emission. Instead, \citet{Glowacki2017} adopted the total hydrogen column density from the X-ray spectrum ($N_{\rm H, X}$) to estimate $T_{\rm S}$ of this target, suggesting that $T_{\rm S}$~$\lesssim$~92~K. They assumed a large \(\int \tau \,d\nu\) considering both narrow and broad components. Thus, this low $T_{\rm S}$ can be taken as the mean $T_{\rm S}$ of most of the infalling gas, particularly the broad component that accounts for a large fraction of the absorption.

The upper limit of $T_{\rm S}$ for a narrow component is also intriguing. Unlike the broad component, which is likely to be an ensemble of multiple components, the narrow gas component can be a single disk rotating around the AGN. Hence, it is reasonable to assume a single linewidth and an upper limit of \ion{H}{1} emission for the narrow component. We thus challenge to constrain $T_{\rm S}$ of the circumnuclear rotating disk using the linewidth of the narrow component ($\sim$65.8~km~s$^{-1}$) and a 3-sigma upper limit of a cleaned JVLA cube with the same beam as the continuum image ($\sigma$~$\sim$~0.97~mJy (=~77.3~K) per beam per channel). The upper limit of the intensity of \ion{H}{1} can be evaluated as in \citet{McNamara1994} by
\begin{equation}
    \int T_{\rm B}~d\nu~\lesssim~3 \times \sigma \times W_{\rm line} \times \left( \frac{W_{\rm chan}}{W_{\rm line}} \right)^{1/2}
\end{equation}
where $T_{\rm B}$ is the brightness temperature in Kelvin, $\sigma$ is the noise level in Kelvin computed in the channel width $W_{\rm chan}$, and $W_{\rm line}$ is the expected linewidth in km~s$^{-1}$.

Assuming that the \ion{H}{1} gas is optically thin ($\tau$~$\ll$~1), its column density can be approximated by the following equation:
\begin{equation}
    N_{\rm HI,emission} = 1.823 \times 10^{18} \int T_{\rm B}~d\nu
\end{equation}
Our assumptions on the emission yield the \ion{H}{1} column density of $N_{\rm HI}$~$<$~1.24~$\times$~10$^{22}$~cm$^{-2}$. Adopting this as the column density of the absorbing gas, the upper limit of the \ion{H}{1} spin temperature can be estimated in Kelvin using the following equation:
\begin{equation}
    N_{\rm HI,absorption} = 1.823 \times 10^{18} ~ T_{\rm S} \int \frac{T_{\rm L}(\nu)}{T_{\rm C}}~d\nu
\end{equation}
where $T_{\rm L}$($\nu$) and $T_{\rm C}$ are the strengths of \ion{H}{1} absorption and continuum emission at 1.4~GHz. Using a maximum dip of 5.5~mJy with an FWHM of 65.8~km~s$^{-1}$ for the narrow absorption component and the continuum flux density of 115~mJy, $T_{\rm S}$ is measured to be $\lesssim$~2041~K. This is much higher than the $T_{\rm S}$ upper limit for the broad component from the X-ray spectrum ($N_{\rm H, X}$) by \citet{Glowacki2017}. This implies the potential presence of a warm neutral medium. Indeed, it has been suggested that high-energy photons emitted by the AGN can excite the surrounding medium in the vicinity of the AGN \citep[e.g.,][]{Bahcall1969, Maloney1996}. Supporting evidence has also been reported in some studies \citep[e.g.,][]{Holt2006, Struve2010}, which found $T_{\rm S}$ of a few times 1000~K near the AGN. This suggests that the accreting gas may not always be ``cold'' and some warm medium can also fuel the AGN.

\subsection{The configuration of the circumnuclear multi-phase gas} \label{subsec:gas_position}

\noindent As shown in Figure~\ref{fig:ALMA_CO_maps2}(a), the radio continuums in kpc-scale observed by JVLA 1.4~GHz and ALMA 103~GHz are unresolved. The observed bipolar jet in pc-scale (Figure~\ref{fig:KaVA_image}) makes it more likely that the true continuum size is very small, which is responsible for producing the observed \ion{H}{1} absorption lines as a background continuum source. As shown in Figure~\ref{fig:ALMA_CO_maps2}(b), none of the detected CO clumps coincide with the BCG center where the AGN resides in. Although Cloud A-2 is closest to the central AGN among all CO clouds, it still barely overlaps with the continuum source in space. In addition, the peak of A-2 spectrum does not match with the peak of either \ion{H}{1} component, even though they span almost the same velocity range (see Figure~\ref{fig:HICO_spectra}). Therefore, it is more natural to think that the central component of the absorbing \ion{H}{1} gas is located deeper inside the BCG than the molecular gas traced by CO emission which is inferred to be yet unsettled in the galactic center.

Meanwhile, as shown in Figure~\ref{fig:HICO_spectra}, the CN absorption \citep{Rose2019b} has an important property in common with the \ion{H}{1} absorption in the sense that it clearly shows two components with one at the BCG center. This suggests that the absorbing CN gas can be distributed similarly to \ion{H}{1}, that is, some as part of the rotating disk around the AGN and some as gas clumps approaching the
center.

\begin{figure*}[ht!]
    \epsscale{0.88}
    \plotone{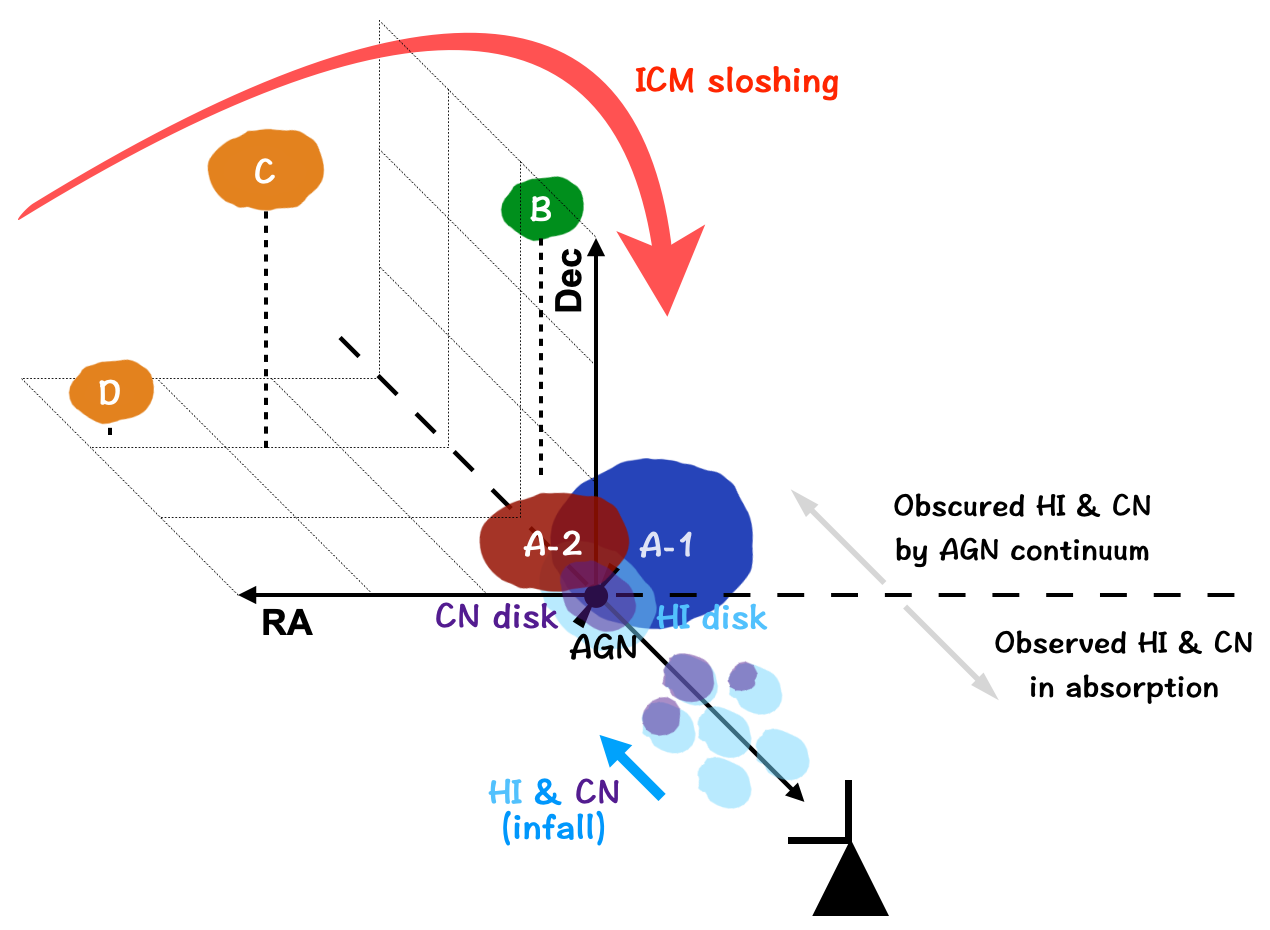}
    \caption{Three-dimensional schematic illustration of multi-phase circumnuclear gases around the AGN in A1644-S. The physical size of each element is not scaled. The black symbol at the center shows the AGN with pc-scale bipolar jets. Both \ion{H}{1} (light blue) and CN (purple) detected as absorption have two distinct components: the central velocity component that is likely part of the rotating disk around the AGN and the redshifted component, which must be the ensemble of infalling clumps on the way to the central SMBH. Molecular gas that has been detected as several CO clumps is thought to be forming along the X-ray sloshing (red arrow) through efficient cooling of the ICM. The locations of CO clouds along the observer’s line-of-sight are inferred from the radial velocities.\\}
    \label{fig:Schematic}
\end{figure*}

These redshifted \ion{H}{1} and CN absorption components can be either radially infalling or non-circularly moving to the AGN, which is not clearly distinguished based on our radio data alone. On the other hand, the H$\alpha$ filaments around the central region of A1644's BCG previously reported are quite intriguing. As shown by \citet[see their Figure~7]{McDonald2012}, both the morphology and the velocity gradient along the filaments are suggestive that the gas is streaming into the center from both sides - front and back. Therefore we favor that the \ion{H}{1} and CN absorbing gases are likely to be part of the radially falling gas toward the center.

Although both \ion{H}{1} and CN have two distinct components, the group velocities of the redshifted part differ by $\sim$100~km~s$^{-1}$ (\ion{H}{1} $\sim$95~km~s$^{-1}$ vs. CN $\sim$200~km~s$^{-1}$ redshifted with respect to the BCG center). If the redshifted components are indeed part of the infalling ISM, this velocity difference may imply that the clumps closer to the center (i.e., high velocity ones with respect to the center) have a higher fraction of CN to \ion{H}{1}. This is also supported by the fact that the formation of CN molecules requires the dissociation of dense molecules such as HCN by high energy \citep[e.g.,][]{Boger2005, Wilson2018}, which is more feasible in the inner part of a galaxy. 

Figure~\ref{fig:Schematic} illustrates a three-dimensional configuration of all observable circumnuclear mediums around the central AGN in A1644-S. The AGN with the pc-scale bipolar jets is shown by the black symbol. Both \ion{H}{1} (light blue) and CN (purple) detected as absorption have two distinct components -- the central component that is likely part of the rotating disk around the AGN and the redshifted component which must be the ensemble of infalling clumps on the way to the SMBH. Meanwhile, CO molecules are thought to be originated from different part compared to the \ion{H}{1} and CN as briefly mentioned above. The arc shape distribution of CO clouds is more suggestive of the formation along the X-ray sloshing (red arrow) through efficient cooling of the ICM.

It remains unclear whether or not there is a CO detection associated with the central \ion{H}{1} and CN components. In fact, we cannot rule out the possibility of CO gas near the central AGN. CO is generally more abundant than CN \citep[e.g.,][]{Meier2015, Wilson2018}, and all seven BCGs with CN absorption studied by \citet{Rose2019b} also show CO absorption except for A1644-S. One possibility is that CO absorption and emission in the center are canceled by each other in this particular target. Alternatively, CO may be visible through higher $J$ transitions. In fact, two out of six BCGs from \citet{Rose2019b} with both absorption lines are detected in CO~(2--1) without CO~(1--0) detection. In the case of A1644-S, no higher $J$ transitions of CO data are available, and the presence of CO gas in its center remains a question to be answered in the future. If CO is detected in the center, either as absorption or emission, its characteristics will provide key information for understanding the connection of the central \ion{H}{1}/CN gas with the off-centered CO and the ICM.

\subsection{The age of the AGN jet} \label{subsec:jet_age}

\noindent The time at which cooling gas flow begins to form until it makes an impact on the AGN activity is key information for understanding the link between ICM cooling and the evolution of BCGs. The assumption that the jet is continuous and symmetric bipolar allows us to estimate how long ago it was launched from the AGN. First, we can infer the possible ranges of viewing angles and bulk velocities using the brightness ratio of the jet and counter jet as follows \citep{Giovannini1994}:
\begin{equation}
    R \equiv I_{\rm Jet}/I_{\rm CounterJet} = (\frac{1+\beta cos\theta}{1-\beta cos\theta})^{2-\alpha}
\end{equation}
where $\beta$ is the jet velocity in light speed units, $\theta$ is the jet viewing angle, and $\alpha$ is the spectral index of the jet. For $\alpha$, we adopt $-$1.0 for this target based on our KVN monitoring program at 22 and 43~GHz \citep{Rose2022}. The SE component is brighter than the NW component, indicating that the NW component is the counter jet, which is the one launched in the opposite direction to an observer. This information yields $R$ of 3.07 and $\beta cos\theta$ of 0.185. This implies a minimum bulk velocity ($\beta_{\rm min}$) of 0.185$c$ when $\theta$ is 0\,$^{\circ}$ (i.e., the jet is perfectly aligned with the direction of our line-of-sight) and the maximum viewing angle ($\theta_{\rm max}$) of 79.3\,$^{\circ}$ when the jet speed approaches the speed of light.

Using the estimated $\beta cos\theta$, we can constrain the deprojected jet size, which can then be used to measure the feasible kinematic age of the jet. Based on our estimates (0.0185~$\lesssim$~$\beta$~$\lesssim$~1), we measured the age for three cases: 0.2$c$, 0.5$c$, and 1$c$. Combining this with the jet extent in the sky of 5~pc, the deprojected jet extents of one side can be 6.6~pc, 2.7~pc, and 2.6~pc. Using the speed of light (0.307~pc~year$^{-1}$), the jet kinematic ages were 107, 18, and 9 years, respectively. This extremely young age of the jet (i.e., a few tens to hundreds of years) is also consistent with the fact that the central continuum source is found to be compact in all non-VLBI data including JVLA 1.4~GHz and ALMA 103~GHz data. This suggests that AGN activity could have been (re)powered quite recently.

\bigskip

\section{Summary} \label{sec:summary}

\noindent In this study, we have reported the ambient medium properties surrounding an AGN located in the center of a cool-core cluster, A1644-S. Our target is a textbook example of the process of cluster merging, where the ICM is thought to have been condensing since the interaction with the northern subcluster approximately 700~Myr ago \citep[][]{Johnson2010}. The main goal of this study is to investigate the origin of CO gas in the BCG and the role of atomic and molecular gases in (re)powering the AGN hosted by the BCG. The important results of our study can be summarized as follows:

\begin{itemize}
    \item The presence of cold gas around the AGN has been confirmed by $^{12}$CO~(1--0) emission and \ion{H}{1} and CN absorption. We find that some of the \ion{H}{1} gas is potentially warm ($\lesssim$~2000~K). 
    
    \item Our high angular resolution KaVA observation enables us to resolve the central source of the BCG in parsec-scale, which reveals a bipolar jet. Together with the fact that there is no hint of more extended radio jets in the center, we suggest the recent resurrection of the central AGN in the past few tens or hundreds of years, which can be the result of cool gas accretion.
    
    \item The arc-shape distribution of $^{12}$CO~(1--0) clouds suggests a connection with hot gas sloshing seen in X-ray from this target. This leads us to the conclusion that hot ICM condensation is a more likely origin of cold molecular gas in this target. This is also supported by the unperturbed stellar morphology of the BCG, which rules out the possibility of molecular gas being tidally induced.
\end{itemize}

\vspace{2mm}
\facilities{Atacama Large Millimeter/submillimeter Array (ALMA), Karl G. Jansky Very Large Array (JVLA), KVN and VERA Array (KaVA)}

\software{Astronomical Image Processing System \citep[AIPS;][]{Greisen2003},  
          Difference Mapping \citep[DIFMAP;][]{Shepherd1997}, 
          Common Astronomy Software Applications \citep[CASA;][]{McMullin2007},
          Cube Analysis and Rendering Tool for Astronomy \citep[CARTA;][]{Comrie2020},
          Groningen Image Processing System \citep[GIPSY;][]{vanderHulst1992}
          }

\acknowledgments

\noindent We thank the anonymous referee for his/her constructive comments and suggestions that helped to improve the paper. This work is based in part on observations made with the KaVA, which is operated by the Korea Astronomy and Space Science Institute and the National Astronomical Observatory of Japan.
This study makes use of the following ALMA data: ADS/JAO.ALMA\#2017.1.00629.S. ALMA is a partnership of ESO (representing its member states), NSF (USA) and NINS (Japan), together with NRC (Canada), MOST and ASIAA (Taiwan), and KASI (Republic of Korea), in cooperation with the Republic of Chile. The Joint ALMA Observatory is operated by ESO, AUI/NRAO, and NAOJ.
The National Radio Astronomy Observatory is a facility of the National Science Foundation, operated under cooperative agreement by Associated Universities, Inc.
Support for this work was provided by the National Research Foundation of Korea (NRF), grant Nos. 2018R1D1A1B07048314 and 2022R1A2C100298211. J.B. acknowledges support by the Graduate School of YONSEI University Research Scholarship Grants in 2019. T.R. thanks the Waterloo Centre for Astrophysics and generous funding to Brian McNamara from the Canadian Space Agency and the National Science and Engineering Research Council of Canada. J.W.K. acknowledges support from the NRF (grant No. 2019R1C1C1002796) funded by the Korean government (MSIT).

\bibliography{sample63}{}
\bibliographystyle{aasjournal}



\end{document}